\documentclass{ws-procs9x6}

\begin{document}

\title{On the practical interest of one-body overlap functions}

\author{Jean-Marc Sparenberg}
\address{TRIUMF Theory Group, 4004 Wesbrook Mall, Vancouver, BC, \\
Canada V6T 2A3, E-mail: jmspar@triumf.ca}  

\author{Byron Jennings}
\address{TRIUMF Theory Group, 4004 Wesbrook Mall, Vancouver, BC, \\
Canada V6T 2A3, E-mail: jennings@triumf.ca}  
\maketitle

One-body (or few-body) functions extracted from full many-body
microscopic wave functions provide a simplified and intuitive insight
into the physics of complex nuclear structure.
They also provide a detailed comparison tool between nuclear models,
which can otherwise only be compared through the quality of their
fit of a few experimental results (spectra, decay widths, etc.).
Since particular one-body functions are also known\cite{escher:02}
to be solutions of Schr\"odinger equations with simple potentials,
they can be used to deduce such potential models from microscopic calculations
by inversion of the Schr\"odinger equation.
In the present work, we deduce the nuclear mean field in $^8$B
from microscopic models;
in the future, we plan to generalize this formalism to one-cluster functions,
which would allow us to deduce nucleus-nucleus potentials
from nuclear cluster models
and baryon-baryon potentials from quark cluster models.

There is however a basic problem for the use of one-body functions:
two alternative functions appear in different theoretical frameworks
of nuclear physics\cite{escher:02} 
and it is not clear which one of them (if any) has the most physical content.
On the one hand, the particle-hole Green-function many-body formalism 
suggests to use the one-body overlap function $\phi(\mathbf{r})$,
whereas the cluster model suggests to use the auxiliary function
$\bar\phi(\mathbf{r})=N^{-1/2}_\mathrm{p}\phi(\mathbf{r})$,
where $N_\mathrm{p}$ is the particle-state norm operator.
In the following, we present first attempts to distinguish between these two
functions.

We have studied the possible difference between
$\phi(\mathbf{r})$ and $\bar\phi(\mathbf{r})$ with respect to the properties
of one-particle decay\cite{al-khalili:03}.
A reduction of the many-body formalism to a one-body formalism for that case
shows that both functions, because of their identical behavior outside the
nuclear range, lead to identical decay widths and hence cannot be distinguished
from one another.
However, we have shown that these two functions lead to different
definitions and possibly to different values for the spectroscopic factor.

\begin{figure}[ht]
\centerline{\epsfxsize=2.1in\epsfbox{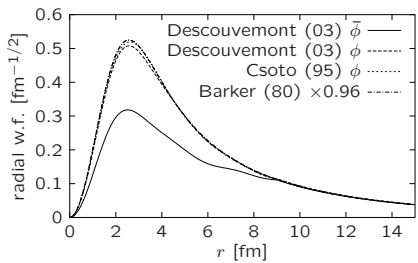}}   
\caption{Radial part of the one-nucleon overlap functions $\phi(\mathbf{r})$
and of the auxiliary functions $\bar\phi(\mathbf{r})$ deduced from
microscopic cluster models, compared with a phenomenological wave function,
for the dominant component of the $^8$B ground state. \label{compwf}}
\end{figure}

More recently, we have compared two microscopic cluster
models\cite{csoto:95,descouvemont:94} with one another and with a
phenomenological potential model\cite{barker:80},
for the $^7$Be($3/2^-$) + p component
of the $^8$B ground state with spin 2 and angular momentum 1.
The corresponding functions are shown in Figure~\ref{compwf}:
the microscopic $\phi(\mathbf{r})$'s are close to one another and to the
phenomenological wave function, which tends to prove that
they have a strong physical meaning.
Moreover, $\bar\phi(\mathbf{r})$ is significantly different from
$\phi(\mathbf{r})$ for the model of Ref.\ \refcite{descouvemont:94}
($\bar\phi(\mathbf{r})$ is not available for the model of Ref.\
\refcite{csoto:95}),
which suggests that $\bar\phi(\mathbf{r})$ does not have such a strong physical
content.
This conclusion is however preliminary since the microscopic functions
from Ref.\ \refcite{descouvemont:94} display a small oscillation
between 5 and 9 fm,
which shows that the numerical convergence of this calculation is not perfect.


\begin{thebibliography}{0}

\bibitem{escher:02}
J.\ Escher and B.~K.\ Jennings,
\newblock {\em Phys.\ Rev.\ C} {\bf 66}, 034313 (2002).

\bibitem{al-khalili:03}
J.\ Al-Khalili, C.\ Barbieri, J.\ Escher, B.~K.\ Jennings, and J.-M.\
  Sparenberg,
\newblock {\em Phys.\ Rev.\ C} {\bf 68}, 024314 (2003).

\bibitem{csoto:95}
A.\ Cs{\'o}t{\'o}, K.\ Langanke, S.~E.\ Koonin, and T.~D.\ Shoppa,
\newblock {\em Phys.\ Rev.\ C} {\bf 52}, 1130 (1995).

\bibitem{descouvemont:94}
P.\ Descouvemont and D.\ Baye,
\newblock {\em Nucl.\ Phys.} {\bf A567}, 341 (1994),
\newblock updated (private communication, 2003).

\bibitem{barker:80}
F.~C.\ Barker,
\newblock {\em Aust.\ J.\ Phys.} {\bf 33}, 177 (1980).

\end{thebibliography}
\end{document}